\documentclass[3p,times,twocolumn]{elsarticle}
 \biboptions{comma,sort&compress}
\usepackage[colorlinks=true,linkcolor=black]{hyperref}

\usepackage{graphicx}
\usepackage{mathtools}
\usepackage{color}
\newcommand{\ep}{\epsilon}
\newcommand{\nn}{\nonumber}
\newcommand{\be}{\begin{equation}}
\newcommand{\ee}{\end{equation}}
\usepackage{here}
\usepackage{ecrc}

\volume{00}
\firstpage{1}
\journalname{Nuclear and Particle Physics Proceedings}
\runauth{Escudero Pedrosa, Llanes-Estrada, Oller and Salas-Bern\'ardez}

\jid{nppp}
\jnltitlelogo{Nuclear and Particle Physics Proceedings}
\usepackage{amssymb}

\begin{document}

\begin{frontmatter}

\title{Assessment of systematic theory uncertainties in IAM unitarization}
 \cortext[cor0]{Communicated to the 23$^{\rm rd}$ International Conference in Quantum Chromodynamics (QCD2020),  Oct. 27$^{\rm th}$--30$^{\rm th}$ 2020, convened online.}
 \author[label1]{Juan Escudero-Pedrosa}
 \author[label1]{Felipe J. Llanes-Estrada\fnref{fn1}}
   \fntext[fn1]{Speaker. E-mail {\tt fllanes@fis.ucm.es}}
\address[label1]{Universidad Complutense de Madrid, Depto.  F\'isica Te\'orica and IPARCOS, 28040 Madrid, Spain.}
\address[label3]{Universidad de Murcia, Departamento de F\'isica, E-30071 Murcia, Spain}
 \author[label3]{Jos\'e Antonio Oller}
 \author[label1]{Alexandre Salas-Bern\' ardez}

\pagestyle{myheadings}
\markright{ }
\begin{abstract}
Effective Field Theories (EFTs) for Goldstone Boson scattering at a low order allow the computation of near--threshold observables in terms of a few coefficients arranged by a counting. As a matter of principle they should make sense up to an energy scale $E\sim 4\pi F$ but the expansion in powers of momentum violates exact elastic unitarity and renders the derivative expansion unreliable at much lower energies.
If new--physics deviations from the Standard Model are found and encoded in low-energy coefficients, perhaps at the LHC, it will be profitable to extend the reach of the EFT to regimes where partial waves are saturating unitarity. The methods known in hadron physics as ``Unitarized Chiral Perturbation Theory'' extend the EFT up to its nominal reach or up to the first new physics resonance or structure (if found below that energy reach) in the partial wave amplitude, but they usually have unknown uncertainties. 
We recapitulate our analysis of the systematic theory uncertainties of the well known Inverse Amplitude Method (IAM).
\end{abstract}
\begin{keyword}  
Theoretical uncertainties \sep Unitarization \sep Effective Theory \sep Inverse Amplitude Medhod 

\end{keyword}

\end{frontmatter}
\section{The Inverse Amplitude Method}

Effective Field Theories can be used to expand a scattering partial wave amplitude
with terms of order $s$, $s^2\dots s^{n+1}\dots$
\begin{equation} \label{EFTExp}
    t\simeq t_0+t_1+\dots
\end{equation}

\begin{figure}[h]\hspace{-0.3cm}
\centering
\includegraphics[width=1.05\columnwidth]{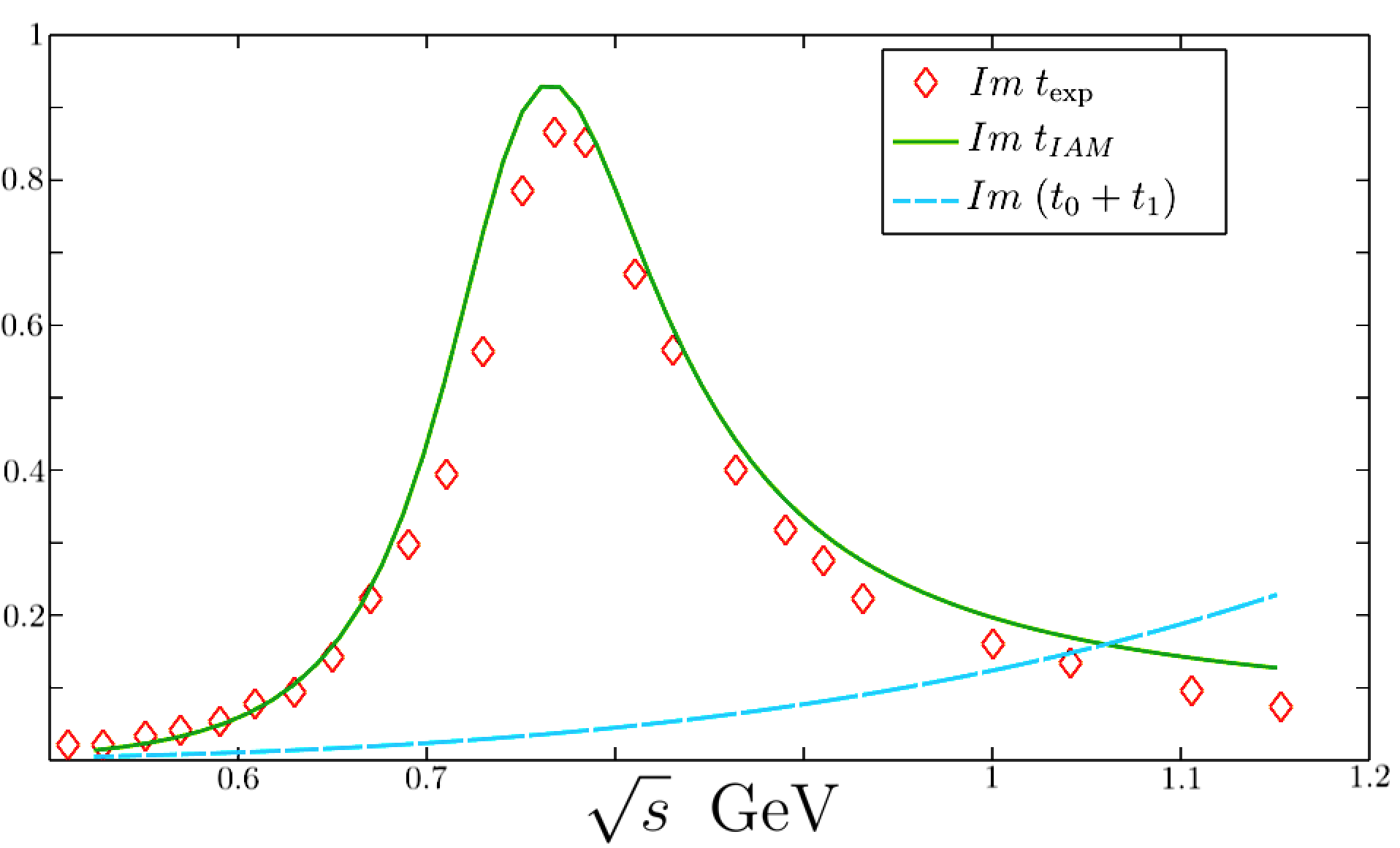}
\caption{\label{fig:rho} {\bf Lowest line}: The perturbative Eq.~(\ref{EFTExp}) fails to track $\pi\pi$ p-wave scattering data. {\bf Upper line}: Implementing Eq.~(\ref{IAM}) reasonably \emph{predicts from threshold data (not from a fit)} the shape and position of the resonance with error under 10\%. This unitarization by the Inverse Amplitude Method extends the validity of the EFT nearer to its nominal expansion convergence at $4\pi f_\pi\simeq 1.2$ GeV).}
\end{figure}

Such expansion satisfies elastic unitarity only in perturbation theory, ${\rm Im}(t_1) = \sigma | t_0|^2$, failing to comply with exact unitarity ${\rm Im}(t) = \sigma | t|^2$. This defficiency of the quasi-Taylor expansion in terms of powers of $s$ with logarithmic modifications spoils the series convergence much before its nominal $4\pi F$ limit scale:  In Fig.~\ref{fig:rho},
the lower discontinuous line (cyan online) fails to track experimental data for $\pi\pi$ scattering data in the $I=1=J$ partial wave featuring the well-known $\rho$ resonance. 

Several methods of unitarization have been devised to extend the EFT's reach, 
but analysis of their systematic theory error are scarce. 
Here we analyse the Inverse Amplitude Method~\cite{Truong:1988zp,2,3,Truong:2010wa,Oller:2020guq} that transforms the perturbative expansion of Eq.~(\ref{EFTExp}) into the following simple form
  \begin{equation} \label{IAM}
 t_{IAM}\equiv \frac{t_0^2}{(t_0-t_1)} \label{usualIAM}     \,.
  \end{equation}

\begin{figure}[htb]
\includegraphics[width=0.85\columnwidth]{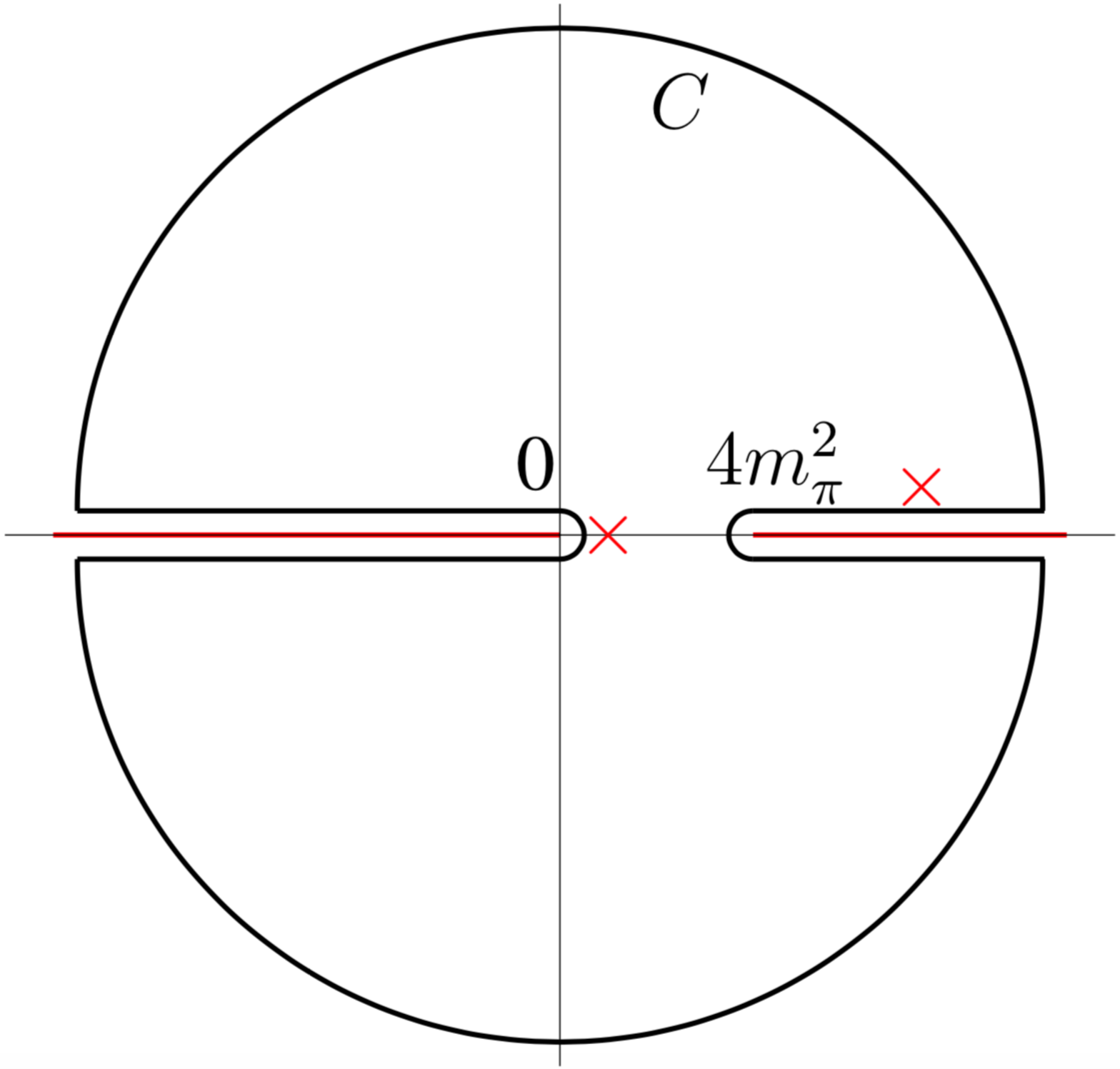}
\caption{
Cauchy's theorem applied within the depicted contour to the auxiliary function of Eq.~(\ref{def:inverse}) leads to Eq.~(\ref{IAM}).
The contour $C$ deforms around the two left (``interaction'') and right (``unitarity'' ) cuts in the complex-$s$ plane.  The two saltires additionally represent the $n$-th order pole at $z=\epsilon$ (where the dispersion relation is subtracted) and the simple pole at $z=s+i\epsilon$ (with $s>4m^2$) coming from the denominators in Eq.~(\ref{disprel}).}
\label{fig:largenenough}
\end{figure}

This is a simple algebraic formula deduced by writing down a dispersion relation for the function 
  \begin{equation}
  G(s)\equiv \frac{t_0(s)^2}{t(s)} \label{def:inverse}
  \end{equation}
from Cauchy's theorem applied to the contour in Fig.~\ref{fig:largenenough}. 

The dispersion relation, appropriately subtracted so convergence for real $|s|\to \infty$ is guaranteed (the partial-wave's diverging exponential factor~\cite{Llanes-Estrada:2019ktp} for complex $s$ is here superconverging for the inverse amplitude), reads
    \begin{align}
  G(s)&= G(\ep)+ G'(\ep) (s-\ep)+\frac{1}{2}G''(\ep)(s-\ep)^2+PC(G)+
   \nonumber\\ &+\frac{(s-\ep)^3}{\pi}\int_{LC}ds'\frac{\text{Im}\,G(s')}{(s'-\ep)^3(s'-s)}+\nn\\&+\frac{(s-\ep)^3}{\pi}\int_{RC}ds'\frac{\text{Im}\,G(s')}{(s'-\ep)^3(s'-s)}
   \label{disprel}\; .
  \end{align}
Eq.~(\ref{IAM}) is obtained by neglecting possible poles of the inverse amplitude ($PC$), approximating with the EFT not only the subtraction constants $G^{(n)}(\epsilon)$ but also the left cut,  even for $s$ beyond its nominal reach, and neglecting all inelastic channels, 
  \begin{align} 
  G(s)&\equiv\frac{t_0(s)^2}{t(s)}=t_0(s)-t_1(s)+\nn\\
  &+\frac{s^3}{\pi}\int_{LC}ds'\frac{\text{Im}\,G(s')+\text{Im}\,t_1(s')}{s'^3(s'-s)}\;.  \label{intermediateG}
  \end{align}
One can also obtain Eq.~(\ref{IAM}) on the real axis by simply derivatively expanding $1/t$ in the EFT. 
But the dispersive derivation has the advantages of being valid in the complex plane so one can look for resonances there, and allowing us to analyse the quality of each of the approximations.
We can quantify the uncertainty introduced by defining an IAM relative separation from the exact partial wave amplitude as 
 \begin{equation} \label{defDelta}
      \Delta(s) = \left(\frac{t_{IAM}(s)-t(s)}{t_{IAM}(s)}\right)\ .
 \end{equation}
 
For new physics applications, in which one would try to predict the mass of a new particle or resonance~\cite{Espriu:2015mya,Delgado:2014dxa,Dobado:2019fxe}
from the measured low-energy coefficients of the EFT, we are most interested in the uncertainty in the mass of that resonance. To estimate  it, we need the uncertainty in the position of the pole at the resonance's mass $s_R$
(with $\sigma=\sqrt{1-4m^2/s}\simeq 1$ the phase space factor),
\be
t_0(s_R)-t_1(s_R)+i\sigma(s_R) t_0^2(s_R)=-\Delta(s_R) G(s_R)\;. \label{corrpoleposition}
\ee

\section{Uncertainty estimate from approximating the LC}

From the definition in Eq.~(\ref{defDelta}) and the dispersion relation in Eq.~(\ref{intermediateG}) we can, for example, isolate the part of the uncertainty that stems from approximating the left cut in perturbation theory, ${\rm Im}\, G \simeq - {\rm Im}\,t_1$, given by 
\begin{equation}\! \! \!
 \Delta(s) G(s)\equiv\Big(\frac{t_{IAM}-t}{t_{IAM}}\Big)\frac{t_0^2}{t}=\frac{s^3}{\pi}\int_{LC}ds'\frac{\text{Im}(G+t_1)}{{s'}^3(s'-s)}\label{IAMerr}
\end{equation}
(other uncertainties being similarly treated~\cite{Salas-Bernardez:2020hua}).

The analysis of this approximation over the left cut is organized by dividing it into three pieces, $I_i\ , i=1,2,3$ 
\begin{align}
\frac{s^3}{\pi}\int_{LC}dz&\frac{\text{Im}\,(G+t_1)}{z^3(z-s)}\equiv\nonumber\\
&\equiv I_1[G+t_1]+I_2[G+t_1]+I_3[G+t_1]\;. \label{threepieces}
\end{align}
We have performed the splitting at the points $|\lambda^2| = 0.47~ {\rm GeV}^2$ (below which we can rely on the EFT)  and $|\Lambda^2| = 2~ {\rm GeV}^2 $  above which we start using the high-energy asymptotic results. Our computation for this high-energy $I_1$  is basically based on the Sugawara-Kanazawa theorem \cite{Sugawara:1961zz}, allowing us to relate the asymptotic behavior on the left cut with that on the physical Right Cut, where Regge theory can be applied in hadron physics.
(In Beyond the Standard Model Physics, a careful analysis would be necessary to see whether all the assumptions needed for this calculation are satisfied, if and when  information thereon starts unfolding). 

The low-$|s|$ contribution $I_3$ lies within the EFT range of applicability so it is easier to constrain, as we know that ${\rm Im}(G+t_1)$ is NNLO there, and thus of order $s^3$.

We find, for $s$ over the right physical cut (in the resonance region), with $k'\sim 0.15$ for the $\rho$-meson channel and $k\sim 1/(4\pi f_\pi)$ in ChPT, 
\begin{eqnarray}
|I_1[G](s,\Lambda,\infty)|&\leq& \frac{k' s^2}{\pi}\log\left( 1+ \frac{s}{\Lambda^2}\right) \\ 
\label{IG1}
|I_3[G+t_1](s,\lambda)|
&\simeq&
 \frac{ks^3}{\pi}\log \Big(1+
 \frac{\lambda^2}{s}
 \Big)\; .
\end{eqnarray}

The intermediate-momentum integral $I_2(G+t_1)$ is most difficult to estimate, so we have adopted three independent schemes.

In the first, we assume ${\rm Im}\,(G+t_1)\leq {\rm Im}\,(t_1)$ which is valid when $G$ and $-t_1$ have similar imaginary parts (remember, this is the case at low momentum), which leads to $I_2\leq 6\%$. However, for the $I=1=J$ channel, this is not a good approximation (see Fig.~\ref{fig:toosmall} below).

A second estimate can be produced noting that the derivative with respect to $s$ of the integral in Eq.~\eqref{IAMerr} should be typically dominated by the low-energy region. By integrating its first derivative one can then employ naturalness arguments to estimate the integration constant, that acts as a counterterm,  in terms of  a scale $L^2\in[2,4]~\text{~GeV}^2$. This procedure yields 
\begin{align}
\label{201024.3}
I_2[G+t_1](s,\Lambda,\lambda)&=\frac{s^3 k}{\pi}\log\frac{s+\lambda^2}{L^2}~,
\end{align}
that numerically evaluates to $I_2\sim 2.4\%$.

Finally, we have made a third order-of-magnitude estimate based on solutions to the GKPY equations~\cite{GarciaMartin:2011cn}
for $\pi\pi$ scattering in the $J=1=I$ partial wave, provided to us by the authors of from~\cite{Pelaez:2019eqa}. These data can give an idea of how the partial wave looks like on the left cut, and the resulting $G$ is depicted in figure~\ref{lcJ}, together with the $t_1$ computed in perturbation theory.
\begin{figure}[h]
\includegraphics[width=\columnwidth]{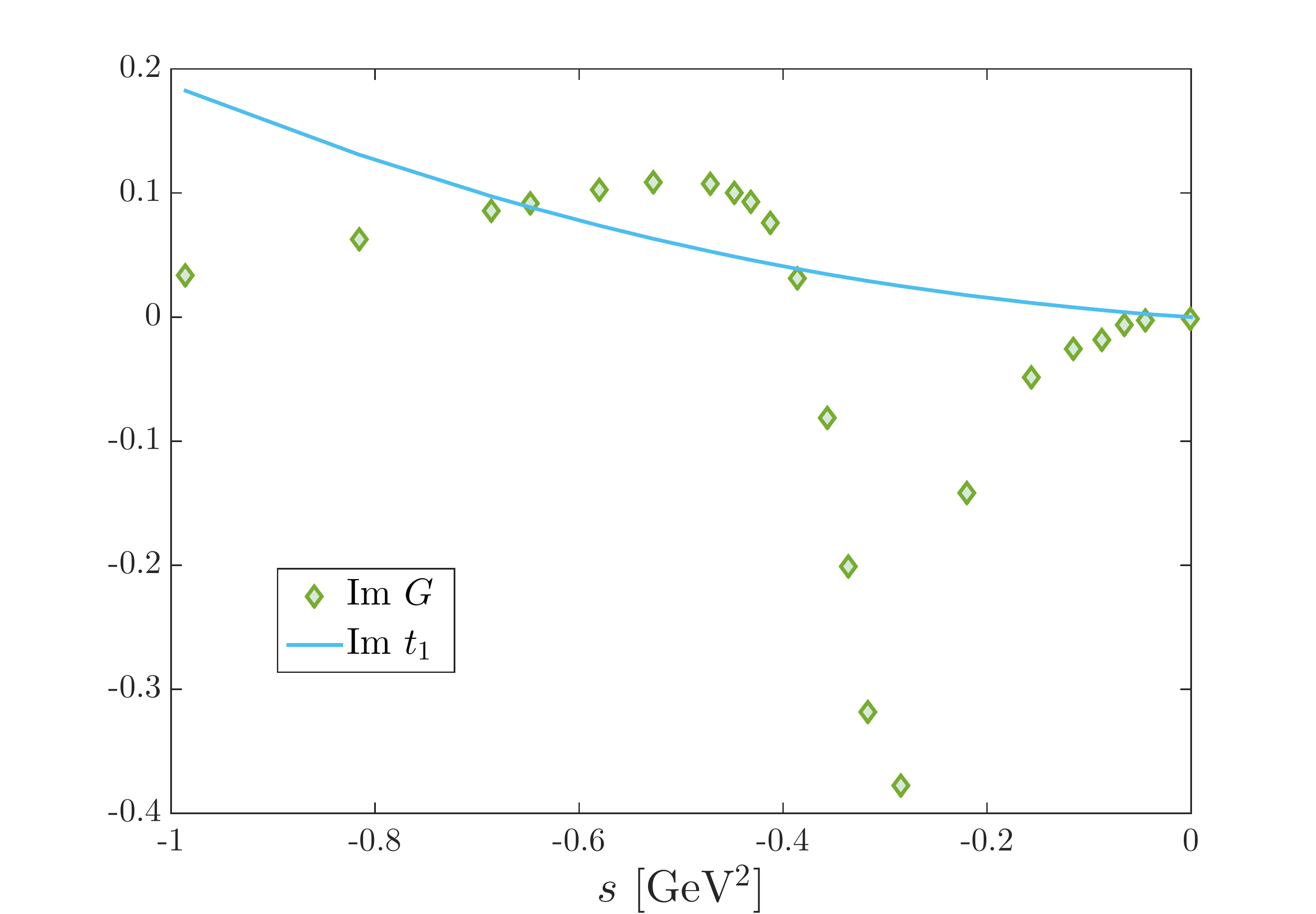}
\caption{
Estimates of the imaginary part of the perturbative $t_1$ in ChPT and auxiliary $G=t_0^2/t$ (from data for $t$ courtesy of Jacobo Ruiz de Elvira~\cite{Pelaez:2019eqa} from solving the GKPY equations),  over the low-energy part of the left cut for $s\in(-1{\rm ~ GeV}^2,0)$.\label{lcJ} 
}
\label{fig:toosmall}
\end{figure}
Both amplitudes are shown up to $s=-1{\rm ~GeV}^2$, and a linear matching is performed to reach $s=-2{\rm ~GeV}^2$ where we start off with the asymptotic behavior. Note that ${\rm Im}(G)$ seems to have an almost-pole (due to an almost-zero of the amplitude $t$) on this channel's left cut. This gives a large contribution to the integral of ${\rm Im}(G+t_1)$: we find $I_2\sim 8\%$. However, this actual value is very sensitive to the degree of cancellation in $t$, which is difficult to pin down along the left cut. In channels without such almost-zero, the uncertainty induced to the IAM would be smaller.

Once the contributions from the three integration intervals are combined, an estimate on the uncertainty in the (``new physics'') resonance pole position has been obtained from Eq.~(\ref{corrpoleposition}).

\newpage
\section{Uncertainty from Adler zeroes and CDD poles}
Poles of the inverse amplitude $1/t\propto G$, that is, zeroes from the actual amplitude $t$, yield additional contributions to Cauchy's theorem denoted as $PC$ in Eq.~(\ref{disprel}). We also divide them into those that appear at low-$s$ near threshold because the EFT derivative terms vanish there (Adler zeroes of the amplitude), and those at higher $s$, perhaps near resonances, that are more influenced by new physics (CDD poles of the inverse amplitude).

The Adler zeroes at $s_A$ have a tiny subpermille level influence on the amplitude in the resonance region, and moreover they can be easily incorporated in a small modification of the IAM itself if one wishes more precision near threshold, by employing instead of Eq.~(\ref{IAM}) the following,
\begin{equation} \label{mIAM0}
  t_{\rm mIAM}\equiv \frac{t_0^2}{t_0-t_1+A_{\rm mIAM}}
\end{equation}
where the denominator has been modified by
\begin{equation}\label{mIAM} \! \! 
  A_{\rm mIAM}\!  =\!  t_1(s_0)\! - \frac{(s_0\! -\!s_A)(s\!-\!s_0)}{s-s_A} (t'_0(s_0)\!-\! t'_1(s_0))
\end{equation}
as has already been known for a time~\cite{GomezNicola:2007qj}.

While the Adler zeroes can be easily seen in the form of the EFT partial wave amplitudes, the CDD poles~\cite{Dyson:1957rgq} are a bit more elusive. If one such pole would fall near a resonance~\cite{Oller:2007xd,Oller:1999me}, it could screen it and a low-$s$ analysis of the amplitude from low-energy measurements could fail to predict it, causing errors of order $O(100\%)$.

To avoid it, we have proposed that the IAM be used in conjunction with a check for zeroes of the amplitude, that from the low-energy EFT can be carried out by examining the condition
\begin{align}
\label{toZeroornottoZero}
t_0(s_C)+\text{Re}t_1(s_C)=0~.
\end{align}
Should this condition reveal a zero of the amplitude at $s_C$, the IAM would be applied to a modified amplitude $t(s)/(s-s_C)$ instead, yielding a modification in the same spirit of Eq.~(\ref{mIAM}), namely
\begin{equation}\label{modIAM2}
t_{\rm IAM} = \frac{t_0^2}{t_0-t_1} \to
\frac{t_0^2}{t_0-t_1+\frac{s}{s-s_c}{\rm Re}(t_1)} \ .
\end{equation}
After this check and eventual modification, that needs to be carried out for each extrapolation of low-energy data with the IAM, the effect of the CDD poles should be controlled.

\section{Inelastic cuts}

The right cut that the IAM takes into account is only the elastic 
Goldstone boson-Goldstone boson cut, {\it e.g.} in $\pi\pi\to \pi\pi$ within chiral perturbation theory, or $\omega_L\omega_L \to \omega_L\omega_L$ in extensions of the electroweak standard model. 

But in the first case, other channels such as $K\bar{K}$, $\eta\eta$, $4\pi$, etc, or in the second case $hh$ or pairs of eventual new particles, can appear. One can write a coupled-channel Inverse Amplitude Method \cite{Oller:1997ng,Oller:1998hw} accounting for the two-body channels, though its dispersive derivation is not straightforward. If this step is not taken, an error in the purely elastic IAM is incurred and we have estimated it in~\cite{Salas-Bernardez:2020hua} for both theories.
To do it, we explicitly show in the imaginary part of the inverse amplitude the piece that is eventually neglected,
\begin{equation}
    \text{Im}\;\frac{1}{t_{\pi\pi}}=-\sigma_{\pi\pi}\Big(1+\frac{\sigma_{K\bar{K}}}{\sigma_{\pi\pi}}\frac{|t_{\pi\pi \to K\bar{K}}|^2}{|t_{\pi\pi\to\pi\pi}|^2}\Big)\;.\label{inelasticim}
\end{equation}
In Chiral Perturbation theory both the ratio of phase spaces (at moderate energies) and the ratio of inelastic to elastic scattering suppress the contribution of the inelastic cut respect to the elastic one in the dispersion relations. Thus, we are relatively sure that below a GeV, the elastic $\pi\pi$ contribution carries the bulk of the right cut.
For the Higgs Effective Field Theory, because the Higgs mass and the $W/Z$ mass are similar (and similarly small respect to $\sqrt{s}$), there is no phase space suppression and we would not know whether the interchannel $\omega_L\omega_L \to hh$ coupling is or not small respect to the elastic one until they are eventually measured.

Additionally, the two-particle right cuts are overlaid by four- (and higher) particle cuts. These are clearly suppressed by the phase space. If the Goldstone boson mass is negligible, these can be written as
\begin{equation}
    \phi_n=\frac{1}{2(4\pi)^{2n-3}}\frac{{s}^{n-2}}{{\Gamma(n})\Gamma(n-1)}
\end{equation}
and otherwise, a numeric computation is necessary. 
The (appropriately dimensionally normalized) ratio of phase spaces is small and controls this uncertainty.

\section{Use of the EFT to approximate the subtraction constants}

The subtraction constants multiplying the $(s-\epsilon)^n$ polynomial terms in Eq.~(\ref{disprel}) are taken at NLO in the EFT. This is a valid approximation because the subtraction point $\epsilon$ is around (below) threshold where the EFT is most accurate. 

Nevertheless, the effect of approximating them (and, partly, the left cut too) to NLO can be checked if/when the NNLO coefficients of the EFT are at hand, and this is just the case in ChPT for $\pi\pi$ scattering~\cite{Bijnens:1998fm}. 

In that case, a formula similar to Eq.~(\ref{IAM}) can be found, incorporating the NNLO order $t_2\propto s^3$, by writing a dispersion relation for
\begin{equation}
G(s)=\frac{t^{2}_0}{t}\simeq \frac{t^{2}_0}{t_{0}+t_1+t_2}\ .
\end{equation}

Without going through the entire dispersive rederivation, we can just expand
\begin{equation}\label{Gtoorder6}
G(s)=\frac{t^{2}_0}{t}\simeq t_{0}-t_{1}-t_{2}+\frac{t^{2}_1}{t_0}\ .
\end{equation}
and then invert to obtain $t^{IAM}$ from the second equality. The size of the two new terms in the denominator, $t_1^2/t_0-t_2$ have been estimated from the known $l_i$ and $r_i$ NLO and NNLO counterterms~\cite{Guo:2009hi}  in Chiral Perturbation Theory (and can be evaluated in Higgs EFT once the corresponding coefficients become known).
The expected size of those new terms becomes the uncertainty estimate on the basic Inverse Amplitude Method if the NNLO contributions are ignored.

\begin{table*}[h]
     \caption{Uncertainty budget for the Inverse Amplitude Method in Hadron Physics. Each entry shows the  scaling with the relevant variables (it is understood that in the resonance region $\sqrt{s}\sim m_\rho \sim M_R$ are similar scales), the order of magnitude of the error they introduce in that resonance region (evaluated in the vector-isovector channel to be specific) and whether the basic method of Eq.~(\ref{IAM}) can be improved to remove that source of uncertainty if needed.}
    \begin{tabular}{lccl} \hline
        Source of uncertainty  & Behavior & Displacement of pole at $\sqrt{s}=m_\rho$  &  Can it be improved?\\ \hline
        Approximate Left Cut & $(\sqrt{s}/(4\pi f_\pi))^4$ & $0.17$ & Partially  \\
        Adler zeroes of $t$   & $(m_\pi/m_\rho)^4$  & $10^{-3}-10^{-4}$ &  Yes: mIAM \\
        CDD poles (zeroes of $t$) at $M_0$ & $M_R^2/M_0^2$ & $0$\,--\,$\mathcal{O}(1)$ & Yes: extract zero before IAM \\
        Inelastic 2-body  & $(\sqrt{s}/(4\pi f_\pi))^4$  &  $10^{-3}$ & Yes: matrix form \\
        Inelastic 4...-body  &$(\sqrt{s}/(4\pi f_\pi))^4$ &$10^{-4}$ &  Partially \\
        $O(p^4)$ truncation  & $(\sqrt{s}/(4\pi f_\pi))^6$ & $10^{-2}$ & Yes: $O(p^6)$ IAM, Eq.~(\ref{Gtoorder6}) \\
    \hline
    \end{tabular}  \label{tab:wrapup}    
\end{table*}

\section{Conclusions}

We have examined the different contributions to the uncertainty budget of one of the best known unitarization
methods that can be applied to a low-energy EFT, the Inverse Amplitude Method of Eq.~(\ref{IAM}). They are collected in Table~\ref{tab:wrapup}.

It is clear that the left-cut approximation is the dominant source of uncertainty. For this $I=1=J$ partial wave, it leads to an {\it a priori} uncertainty of  $O(17\%)$. We know, {\it a posteriori} (Fig.~\ref{fig:rho}) that the $\rho$ is reproduced to 10\%; In the case of Higgs EFT we would at best find a separation from the Standard Model near threshold, so that only the {\it a priori} estimate would be at hand.   

While other authors have worried about systematic uncertainties in the IAM~\cite{Garcia-Garcia:2019oig,Corbett:2015lfa}, we have presented a comprehensive analysis of the status of those uncertainties, with an explicit evaluation in Chiral Perturbation Theory where data on pion-pion scattering is at hand, and provided working guidelines on how the computation can be approached should coefficients of Higgs EFT be found to separate from the Standard Model at the LHC.
While we have concentrated on the $s$-dependence of the scattering amplitude, 
other authors~\cite{Niehus:2020gmf} have also examined the IAM to perform extrapolations in theoretical parameters (saliently, the quark mass) that can be of assistance for exploitation of lattice data. 
The interested reader can find much many additional details in an upcoming publication~\cite{Salas-Bernardez:2020hua}.

\section*{Acknowledgments}

We thank Prof. S. Narison and his Montpellier CNRS team for providing, year after year, a venue for discussion of new results in the strong interactions and QCD, and especially for preparing it fully online under this year's conditions.
We  acknowledge support by EU Horizon 2020 research and innovation programme, STRONG-2020 project, grant agreement No 824093; 
grants FEDER (EU) and MINECO/MICINN (Spain) FPA2016-75654-C2-1-P and -77313-P, PID2019 -108655GB-I00, -106080GB-C21 (Spain) -106080GB-C22; Universidad Complutense de Madrid under research group 910309 and the IPARCOS institute; and the VBSCan COST Action CA16108.


\end{document}